\documentclass[%
 reprint,
 floats,
nofootinbib,
 amsmath,amssymb,
 aps,
 prc
]{revtex4-2}

\usepackage[english]{babel}

\usepackage{
            amsmath,
            hyperref,
            url,
            graphicx,
            color}

\newcommand\unit[1]{~\mathrm{#1}}
\newcommand\MeV[1]{#1\unit{MeV}}

\newcommand\op[1]{#1}
\newcommand\mat[1]{#1}

\newcommand\of[1]{\left(#1\right)}

\newcommand\bra[1]{\left<#1\right|}
\newcommand\ket[1]{\left|#1\right>}

\newcommand\braopket[3]{\left<#1\middle|#2\middle|#3\right>}

\newcommand\hconj[1]{{#1}^\dagger}

\newcommand\id{\mathrm{I}}

\newcommand\isotope[2]{{}^{#1}\mathrm{#2}}
\newcommand\mg[1]{\isotope{#1}{Mg}}

\usepackage{accents}

\begin{document}

\renewcommand{\figurename}{Fig.}
\renewcommand{\figureautorefname}{Fig.}

\title{Nuclear cross sections from low-energy interactions}

\author{J. Bostr\"om}
\affiliation{Division of Mathematical Physics, Department of Physics, LTH, \\ Lund University, PO Box 118, S-22100 Lund, Sweden}
\author{J. Rotureau}
\affiliation{Division of Mathematical Physics, Department of Physics, LTH, \\ Lund University, PO Box 118, S-22100 Lund, Sweden}
\author{B. G. Carlsson}
\affiliation{Division of Mathematical Physics, Department of Physics, LTH, \\ Lund University, PO Box 118, S-22100 Lund, Sweden}
\author{A. Idini} 
\affiliation{Division of Mathematical Physics, Department of Physics, LTH, \\ Lund University, PO Box 118, S-22100 Lund, Sweden}

\begin{abstract}
    We present a method to calculate neutron scattering cross sections for deformed nuclei using many--body wavefunctions described with multiple reference states. Nuclear states are calculated with the generator coordinate method using a low energy effective Hamiltonian. Using these states, a non--local and energy dependent optical potential is consistently constructed, allowing to directly investigate the role of nuclear structure properties in nuclear scattering.
    The case of neutron scattering on $^{24}$Mg is presented.
    The results are compared to experiment and to phenomenological optical potentials at energies below 13 MeV, demonstrating the importance of low--energy collectivity in elastic and non--elastic scattering.
\end{abstract}

\maketitle

\section{Introduction}

Nuclear reactions are one of the fundamental methods used to study and understand atomic nuclei. Exotic nuclei, which must be produced in radioactive ion beam facilities and studied before they decay, are often investigated through reaction processes \cite{Johnson:20,Crawford:24}. Reactions are not only crucial for understanding nuclei but also play a significant role in various astrophysical phenomena, for example stellar burning and nucleosynthesis \cite{Nunes:05,Schatz:22}, as well as reactor physics, where reactions such as neutron induced reactions have practical applications and are important in the quest of generating microscopic data for simulations of new types of reactors \cite{Rochman:25}.

It is extremely challenging to study nuclear reactions using state--of--the--art nuclear structure information in a consistent framework. The complexity of the dynamical processes happening during a scattering process often forces the use of separate and inconsistent models of structure and reaction, frequently relying on phenomenological optical potentials \cite{Koning:03,Pruitt:23}. Optical potentials represent the effective interaction between projectile and target and they are an effective way to decouple internal degrees of freedom and reaction dynamics \cite{Feshbach:58b}. They can be calculated exactly from the Hamiltonian of the many--body problem projected onto the elastic scattering channel, as shown already in \cite{Feshbach:58a}.
Recently, several efforts have been made to calculate cross sections and produce adequate optical potentials consistently using microscopic nuclear structure models from Hamiltonian projection (cf. \cite{Johnson:20,Hebborn:23} and refs. therein).

It is additionally difficult to describe reactions involving deformed nuclei. The symmetry breaking mechanisms that efficiently describe the deformation, impose an additional complexity to both the formalism and the computation of nuclear properties and their application to reactions \cite{Idini:23proc}.

In this manuscript, we propose a novel method for constructing an optical potential for deformed nuclei using microscopic symmetry breaking and restoration calculations. The formalism needed to construct Green's functions and corresponding self--energy from multiple reference states is presented. In particular, a sum rule is introduced that allows to represent a large Fock space, necessary for reaction observables, while applying a truncation to the many-body space relevant for structure properties, reducing computational time.

This method is applicable to a wide range of interactions and for the whole nuclear chart. Here, as a test case, we use it to calculate scattering cross sections of $n+^{24}$Mg.

Our approach builds on previous work utilizing the generator coordinate method (GCM) with an effective low--energy interaction \cite{Ljungberg:22}. This combination is a versatile many--body framework capable of describing both light and superheavy deformed nuclei \cite{Samark:21,Ljungberg:22,Ljungberg:23proc}, and extends the GCM with a procedure that allows it to account for single--particle excitations. The GCM combined with symmetry restoration and carefully chosen collective degrees of freedom is an efficient way of obtaining many-body solutions that can be very close to exact solutions \cite{Bally:21}.

The method allows investigating the role of many--body correlations in scattering, with a particular focus on the effects of low--energy collectivity on the observed elastic scattering cross sections. In \cite{Idini:19} self--consistent Green's functions were used to calculate neutron elastic scattering of $^{16}$O and $^{40}$Ca, noting that the overestimation of elastic scattering cross section was due to lacking correlations and collectivity. This method alleviates this issue, concluding that GCM can reproduce low--energy collectivity important for reaction properties. These results show a promising route towards the systematic construction of microscopic optical potentials for heavy and deformed nuclei.

\section{Method}

The method to construct an optical potential from a microscopic calculation presented in this paper relies on calculating the time--ordered Green's function of the system, and then finding the associated self energy. The unperturbed Green's function $\op{G}_0\of{E}$ represents the propagation of a particle in a fixed external potential $\op{U}^0$, while the dressed propagator $\op{G}\of{E}$ considers the effect of the interaction of the particle with a many--body system. Both can be seen as operators on the single--particle space. The two are related through the self energy $\op{\Sigma}\of{E}$ and the Dyson equation
$\op{G}\of{E} = \op{G}_0\of{E} + \op{G}_0\of{E} \op{\Sigma}\of{E} \op{G}\of{E}$,
which can be solved for $\op{\Sigma}\of{E}$ as
\begin{equation}
    \label{eq:dysoninv}
    \op{\Sigma}\of{E} = \op{G}_0\of{E}^{-1} - \op{G}\of{E}^{-1}.
\end{equation}
The sum of the external potential and the self energy,
$\op{V}\of{E} = \op{U}^0 + \op{\Sigma}\of{E}$,
which is independent of the choice of external potential,
can then be identified with the optical potential experienced by an additional particle scattering on the system \cite{Capuzzi:96}.

Using complete sets of states for the systems with one more and one less particle, the matrix elements of the time--ordered Green's function's can be written in the K\"all\'en--Lehmann representation as
\begin{align}\label{eq:klrep}
    G_{\alpha,\beta}\of{E} = \lim_{\eta\rightarrow 0^+} &\sum_i \frac{\braopket{\Psi_0}{\op{a}_\alpha}{\Psi^+_i}\braopket{\Psi^+_i}{\hconj{\op{a}}_\beta}{\Psi_0}}{E - (E^+_i - E_0 - i\eta)} \notag\\+ &\sum_i \frac{\braopket{\Psi^-_i}{\op{a}_\alpha}{\Psi_0}\braopket{\Psi_0}{\hconj{\op{a}}_\beta}{\Psi^-_i}}{E + (E^-_i - E_0 - i\eta)},
\end{align}
where $\ket{\Psi_0}$ represents the ground state of the system with $A$ particles and energy $E_0$, $\ket{\Psi^\pm_i}$ are a complete set of energy eigenstates of the system with $A\pm1$ particles with energies $E^\pm_i$, $\op{a}_\alpha$ are the annihilation operators of the single--particle basis in $m$--scheme, and the infinitesimal $\eta$ ensures the correct time--ordering.
In the continuum energy region, the sums would become integrals. Due to $\eta$, this contributes a non--Hermitian part to the Green's function. This non--Hermitian part is directly related to the scattering states, and is therefore important in the description of scattering.

Using wavefunctions obtained from a many--body solution, one can calculate the spectroscopic amplitudes
of the $A\pm1$--particle states with respect to
the $A$--particle ground state $\Psi_0$, defined as
$s^+_{i,\alpha}\equiv\braopket{\Psi^+_i}{\hconj{\op{a}}_\alpha}{\Psi_0}$ and $s^-_{i,\alpha} \equiv \braopket{\Psi_0}{\hconj{\op{a}}_\alpha}{\Psi^-_i}$.
The spectroscopic amplitudes represent how well the states $\ket{\Psi^\pm_i}$ are described as
a single particle in the state $\alpha$ added on (or removed from) the $A$--particle ground state $\Psi_0$.
Their absolute squares,
$\left|s\right|^2$,
are called spectroscopic factors.

Since many--body solution methods must truncate the full Fock space, the set of solutions obtained might not be complete, and in this case the K\"all\'en--Lehmann representation \eqref{eq:klrep} cannot be used directly. Here we propose a method to augment the solution set with additional states so that the K\"all\'en--Lehmann representation can still be used and the Green's functions can be calculated.
The idea is that the many--body solution will include the most important contributions from correlated states in the energy range of interest, and that the remaining part can be approximated by an additional mean field.
To find this mean field, we consider two sum rules. First, we have the completeness relation
$\sum_{i,x=\pm} \left(s^x_{i,\alpha}\right)^* s^x_{i,\beta} = \delta_{\alpha,\beta}$, which corresponds to stating that the solutions of the $A\pm1$ system fully spans the space of single--particle and single--hole excitations on the $A$--particle ground state $\ket{\Psi_0}$. Secondly, we consider the expression $(H_A)_{\alpha,\beta} \equiv \bra{\Psi_0} \left\{ \op{a}_\alpha, \left[ \op{H}, \hconj{\op{a}}_{\beta} \right] \right\} \ket{\Psi_0}$, which we can expand and insert the completeness relations $\sum_i \ket{\Psi^\pm_i}\bra{\Psi^\pm_i} = \op{I}^\pm$, where $\op{I}^\pm$ are the identity operators on the spaces of $A\pm1$ particles, to get
\begin{align}
    (H_A)_{\alpha,\beta} &=
    \sum_i (E^+_i - E_0) (s^+_{i,\alpha})^* s^+_{i,\beta} \\&+ \sum_i (E_0 - E^-_i) s^-_{i,\beta} (s^-_{i,\alpha})^*. \notag
\end{align}
Defining $\bar{E}^\pm_{i} = \pm\left(E^\pm_i - E_0\right)$, this can be written as
\begin{equation}
    \sum_{i,x=\pm} (s^x_{i,\alpha})^* \bar{E}^x_{i} s^x_{i,\beta} = \bra{\Psi_0} \left\{ \op{a}_\alpha, \left[ \op{H}, \hconj{\op{a}}_{\beta} \right] \right\} \ket{\Psi_0}.
    \label{eq:ewsr}
\end{equation}
This is related to the spectroscopic sum rule derived for nuclear matter in \cite{Polls:94}. The right hand side is discussed in \cite{Rowe:68}, where it is identified as the matrix elements of a generalized single--particle mean field for the state $\ket{\Psi_0}$, and when $\ket{\Psi_0}$ is a Hartree--Fock state, it reduces to the matrix elements of the mean field Hamiltonian. Its eigenvalues are also identified as the experimentally observable centroids for the total stripping and pick--up strengths of the single--particle states \cite{Rowe:68}.

For an incomplete set of states, these sum rules will in general not be fulfilled
.
However, using the sum rules we can find an estimate of the average contribution of the missing states, and allow us to introduce $N$ additional $A\pm1$--particle states, where $N$ is the number of single--particle states in the basis.
Denoting the spectroscopic amplitudes and energies of the additional states $c_{k,\alpha}$ and $\epsilon_{k}$, they must then satisfy the sum rules in the following way,
\begin{align}
    \sum_{i,x=\pm}(s^x_{i,\alpha})^* s^x_{i,\beta} + \sum^{N}_{k=1}(c_{k,\alpha})^* c_{k,\beta} = \delta_{\alpha,\beta},
    \label{eq:completion1} \\
     \sum_{i,x=\pm}(s^x_{i,\alpha})^* \bar{E}^x_{i} s^x_{i,\beta} + \sum^{N}_{k=1}(c_{k,\alpha})^* \epsilon_{k} c_{k,\beta} = (H_A)_{\alpha,\beta}.
    \label{eq:completion2}
\end{align}
In the case of a rotationally and parity invariant system, the sum rules hold for each spin $J$ and parity $\pi$, and in that case, $N = N^{J\pi}$ is the number of shells with the given spin and parity.

With the number of additional amplitudes and energies equal to the number of single--particle states, $N$, the sum rules determine the additional states uniquely.
This can be seen by rewriting the equations in matrix form. With $\bar{E}_{i j}=\bar{E}_i \delta_{i j}$ and $\epsilon_{i j}=\epsilon_i \delta_{i j}$, the first sum rule then becomes $\hconj{\mat{s}} \mat{s} + \hconj{\mat{c}} \mat{c} = \mat{\id}$, and the second sum rule $\hconj{\mat{s}} \mat{\bar{E}} \mat{s} + \hconj{\mat{c}} \mat{\epsilon} \mat{c} = \mat{H_A}$.
By using the polar decomposition of $\mat{c} = \mat{U} \mat{P}$, where $\mat{U}$ is unitary and $\mat{P}$ is Hermitian, we find $\mat{P}^2 = \mat{\id} - \hconj{\mat{s}} \mat{s}$ using the first rule,
and $\hconj{\mat{U}} \mat{\epsilon} \mat{U} = \mat{P}^{-1}(\mat{H_A} - \hconj{\mat{s}} \mat{\bar{E}} \mat{s})\mat{P}^{-1}$ using the second.
These can be uniquely solved using a matrix square root and as an eigenvalue equation, respectively.

With no many--body states with non--zero $s^x_{i,\alpha}$, the additional spectroscopic amplitudes and energies would be the eigenstates and eigenvalues of $\mat{H_A}$, the generalized mean field.
As the many--body solution converge and come closer to fulfilling the sum rules,
the added $c_{k,\alpha}$ go to zero, and their contribution diminishes.
The added contribution is therefore interpreted as the mean field average of the neglected states.

With these spectroscopic amplitudes and energies, we then construct the Green's function in the K\"all\'en--Lehmann representation using Eq. (\ref{eq:completion1},\ref{eq:completion2}) as,
\begin{equation}\label{eq:kl}
    G_{\alpha,\beta}\of{E} = \sum_{i,x=\pm} \frac{(s^x_{i,\alpha})^* s^x_{i,\beta}}{E-\bar{E}^x_{i} \of{\eta}} + \sum_k\frac{(c_{k,\alpha})^*c_{k,\beta}}{E- \bar{\epsilon}_{k}\of{\eta}},
\end{equation}
where $\bar{E}^\pm_{i} \of{\eta} = \pm\left(E^\pm_{i} - E_0 - i\eta\right)$.
$\bar{\epsilon}_{k}\of{\eta}$ is chosen as $\bar{\epsilon}_{k}\of{\eta} = \epsilon_{k} + i\eta$ for $\epsilon_{k} < E_\mathrm{F}$ and $\bar{\epsilon}_{k}\of{\eta} = \epsilon_{k} - i\eta$ for $\epsilon_{k} > E_\mathrm{F}$, where $E_\mathrm{F}$ can be set appropriately according to the French-McFarlane sum rules \cite{Bostrom:upcoming,suppl} but it can be taken as the Fermi energy $(E^+_0 - E^-_0)/2$ with no difference to the results presented here.

Due to the discrete basis, continuum effects are not included, and so a finite value of $\eta$ is necessary to get nonelastic scattering. This is equivalent to treating the discrete energies as resonances with widths proportional to $\eta$. There are methods for calculating the appropriate imaginary parts of the energies $\bar{E}$, for example using a Berggren basis \cite{Berggren:68}, as done in \cite{Rotureau:17}, however such a method is not employed in this first application of this method. In the results we will use a simpler recipe to set $\eta$ to an average resonance width.

To demonstrate the method, we employ the Hamiltonian described in \cite{Ljungberg:22} $\op{H} = E_0 + \op{\Gamma} + \op{V}$, where $E_0$ is a constant, $\op{\Gamma}$ and $\op{V}$ are the one and two--body components respectively. The form may be derived starting from a general interaction by a normal ordering procedure that approximates the three--body interaction \cite{Lin:24}. In our case, the terms are given by a low--energy effective Hamiltonian that captures the response of an energy density functional to external fields \cite{Ljungberg:22}.

The many--body basis that is used to solve the Hamiltonian consists of Hartree--Fock--Bogoliubov (HFB) vacua varied over a set of generator coordinates. The collective coordinates that generate the GCM basis are the familiar $\beta$, and $\gamma$ for quadrupole deformation and triaxiality, in addition to a variation of the neutron (proton) pairing fields scaled by $g_n$ ($g_p$) factors, and different cranking constraint $j_x$. To also include single--particle excitations, each HFB state $\ket{\Phi (\beta, \gamma, g_n, g_p, j_x)}$ is also excited with a Bogoliubov singles coupled cluster operator with a temperature--like weighting obtaining $\ket{\Phi_x}$ \cite{Ljungberg:22}.
This choice of generator coordinates and reference states accounts for the most important degrees of freedom of single--particle, collective vibrations, rotations and pairing vibrations already within the reference states.
Additionally, to describe states with an odd number of particles, we apply the quasiparticle creation operator $\hconj{\beta}_a$ to each HFB reference state,
$\ket{\Phi^{\pm}_{a,x}} = \hconj{\beta}_a\ket{\Phi_x}$.

These basis states are then projected to good angular momenta and particle numbers to calculate Hamiltonian and overlap matrices. The Hill--Wheeler equation is then constructed and solved, finally obtaining states $\ket{{\Psi^\pm}^{J\pi}_i} = \sum_{a,x,M} h^{J\pi}_{a,x,i,M} P^J_{K,M} P_Z P_{N\pm1} \ket{\Phi^\pm_{a,x}}$, where $\Psi^\pm_i$ denote
states with $A \pm 1$ particles with energy $E^\pm_i$, and $\Psi_i$ denotes
a state with $A$ particles with energy $E_i$. $J$ is the total angular momentum,
$\pi=\pm 1$ is the parity, $i$ the label of the state,
$h^{J\pi}_{a,x,i}$ are coefficients,
and $P^J_{M,K}$, $P_Z$ and $P_N$ are the projection operators for angular momentum, proton number and neutron number respectively.
This method can be used with a wide variety of interactions and is able to cover the whole nuclear chart.
The method described in this paper extends the GCM approach to also be used for scattering calculations.

%
%
%


When calculating using a spherically symmetric single--particle basis,
and since an even--even ground state will have spin 0 and positive parity,
the spectroscopic amplitude is only non--zero when
the spin and parity of the operator $\hconj{a}_\alpha$ matches
the spin and parity of the odd--even state ${\Psi^+}^{J\pi}_i$,
so $J_\alpha = J$ and $\pi_\alpha = \pi$,
and it's only necessary to project the ket \cite{Hakansson:78,Enami:99}.
More information regarding the calculation of the spectroscopic factors can be found in \cite{Bostrom:23proc,Bostrom:upcoming}.

We then construct $G_{\alpha,\beta}\of{E}$ using \eqref{eq:kl} and the GCM solutions. $(G_0)_{\alpha,\beta}\of{E}$ is constructed using the spherical Hartree--Fock solution used to define $\op{\Gamma}$ \cite{Ljungberg:22}, substituting in Eq. (\ref{eq:kl}) $s^x_{i,\alpha}$ and $\bar{E}^\pm_{i} \of{\eta}$ with the spectroscopic amplitudes and energies of the HF excitations, and $c_{k,\alpha}$ with 0.
We can then calculate the optical potential $V_{\alpha,\beta}(E)=U^0_{\alpha,\beta} + \Sigma_{\alpha,\beta}(E)$ using \eqref{eq:dysoninv}, with $\op{U}^0 = \op{\Gamma} - \op{T}$ as the potential part of $\op{\Gamma}$ while $\op{T}$ is the kinetic part.

To improve convergence, a smoothing factor is applied to the potential, as was proposed in \cite{Revai:85} and used in several scattering calculations \cite{shirokov:2021,du:2022}. The factor reduces the effect of basis truncation and the details of its implementation can be found in \cite{suppl}.
With this smoothed potential, the momentum space potential $V^{J\pi}\of{p, p', E}$ is calculated using the momentum space single--particle wavefunctions.
The momentum space Schr\"odinger equation describing the scattering neutron of energy $E_\mathrm{cm}$ in the center of mass frame for a given partial wave is
\begin{align}
    \frac{p^2}{2\mu}u\of{p} + \gamma^3 \int \mathrm{d}p'\,p'^2\,V^{J\pi}\of{\gamma p,\gamma p',\gamma E_\mathrm{cm}} u\of{p'} \notag \\ = E_\mathrm{cm} u\of{p},
    \label{eq:scattering}
\end{align}
where $\gamma \equiv m_1/\mu = 1+1/A$, $\mu=m_1 m_2/(m_1 + m_2)$ is the reduced mass, $m_1$ and $m_2$ are the projectile and target masses \cite{Idini:19}.
The Schr\"odinger equation in the laboratory frame is obtained substituting the reduced mass with the projectile mass, $\gamma$ with 1, and $E_\mathrm{cm}$ with the projectile energy in the laboratory frame $E_p$.

This Schrödinger equation is then solved
using the Lippmann--Schwinger equation in momentum space,
giving the phase shifts for each partial wave, $\delta_{J\pi}$.
The phase shifts are then used to calculate differential
cross sections $\mathrm d\sigma/\mathrm d\Omega$,
as well as integrated elastic, reaction, and total cross sections
$\sigma_E$, and $\sigma_T$,
in the same way as in \cite{Arellano:21}.

\section{Results}

As the first implementation of this method, we have calculated total and elastic neutron scattering cross sections of the characteristically prolate deformed nucleus $\mg{24}$.
The calculation of the nuclear wavefunctions were executed following the framework described in \cite{Ljungberg:22, Bostrom:23proc}. The effective Hamiltonian was created for $^{24}$Mg using the SLy4 Skyrme parameterization \cite{Chabanat:98}
.
SLy4 was previously used to calculate properties of $^{24}$Mg in \cite{Ljungberg:22,Bender:08}. Using this Hamiltonian and basis states, the eigenstates of $\mg{23,24,25}$ were calculated \cite{Ljungberg:22,Bostrom:23proc}.
The convergence parameters are described in \cite{suppl}.
The wavefunctions were then used to calculate the spectroscopic amplitudes of $\mg{23}$ and $\mg{25}$ relative to the ground state of $\mg{24}$ and to construct the corresponding Green's functions \eqref{eq:kl}.

\begin{figure}[t]
    \includegraphics[width=\linewidth]{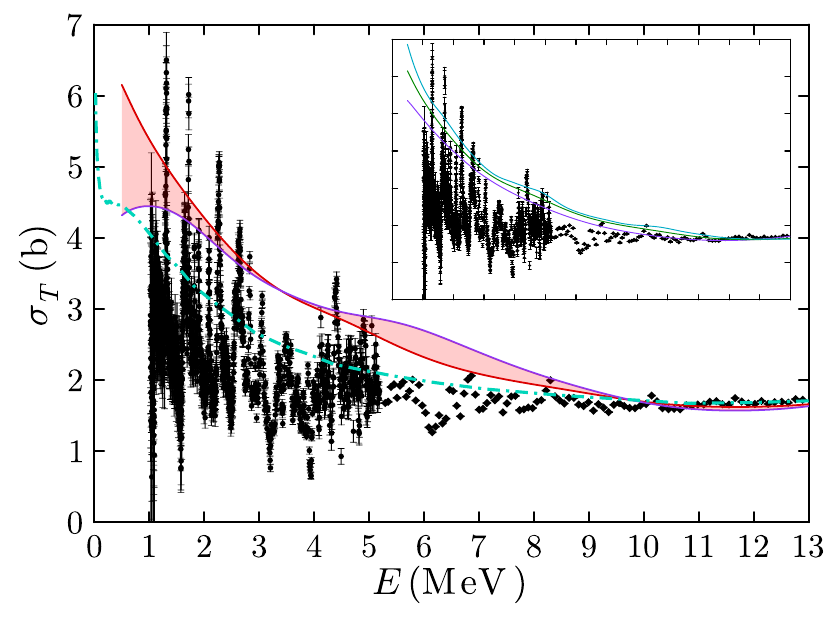}
    \caption{The total scattering cross section of $\mg{24}+n$ in function of the neutron energy in the lab frame. In the main figure, the red and purple lines correspond to the cross sections calculated using the method described in this paper with signature $+i$ and $-i$ respectively and the area between them is shaded red.
        The cyan dashdotted line shows the result using the Koning--Delaroche optical potential. The black circles are experimental cross sections for $\mg{24}$ \cite{Bommer:76}, while the diamonds are natural Magnesium \cite{Abfalterer:01}. In the inset, the results for $+i$ signature and variation of $\eta$ parameters, with $a=6,12,24$ MeV for light blue, green and purple line respectively.}
    \label{fig:total}
\end{figure}

For the scattering calculations, $\eta$ was set to
$\eta = \frac{a}{\pi}\frac{\left(E-E_f\right)^2}{\left(E-E_f\right)^2+b^2}$
with $a=\MeV{12}$ and $b=\MeV{22.36}$ following a common prescription to represent the average resonance widths \cite{Waldecker:11,Brown:81}. A variation of a factor 4 of the $\eta$ parameter obtains scattering results generally close to the provided convergence band, as shown in the inset of Fig. \ref{fig:total}. $a=\MeV{24}$ gives on average $8.8\%$ lower total cross section than $a=\MeV{6}$, much smaller than the factor 4 increase of the average resonance widths represented by $\eta$.
Methods for treating the imaginary part of the Green's function in \eqref{eq:kl} have been developed for example in \cite{Sargsyan:24} using self--consistency or in \cite{Johnson:20,Rotureau:17} using continuum states. Such a consistent approach is not within the scope of this paper, but since the method can be seen to perform well for a wide range of $\eta$, it would not change the conclusions of this paper. 
Neutron scattering on $^{24}$Mg was also calculated in \cite{Sargsyan:24,Kravvaris:24} with single--reference configuration interaction (shell model) prescriptions, making it an interesting case of comparison.

\begin{figure}[t]
    \includegraphics[width=\linewidth]{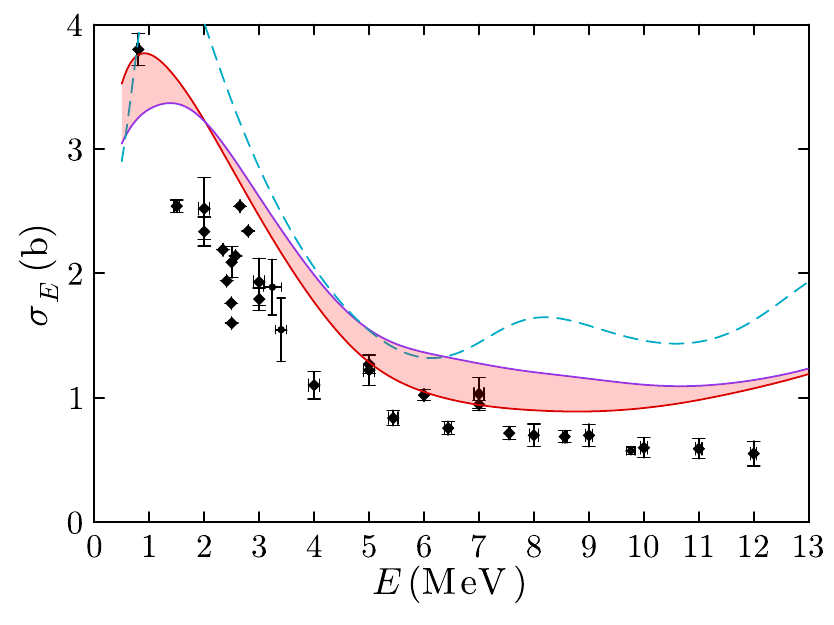}
    \caption{The elastic scattering cross section
        of $\mg{24}+n$ in function of neutron energy in the lab frame.
        The lines and datapoints mean the same as in \autoref{fig:total}.
        Additionally, the integrated cross section calculated with only the completion states is shown as a dashed blue line.
        The experimental cross sections
        for $\mg{24}$ are from \cite{Virdis:81,Schweitzer:78,Frittelli:70},
        while the natural Magnesium cross sections are from
        \cite{Korzh:94,AdelFawzy:85,Korzh:69,Korzh:64,Korzh:63,Thomson:62,Little:46,MacPhail:40}.}
    \label{fig:el}
\end{figure}

\begin{figure}[t]
    \includegraphics[width=\linewidth]{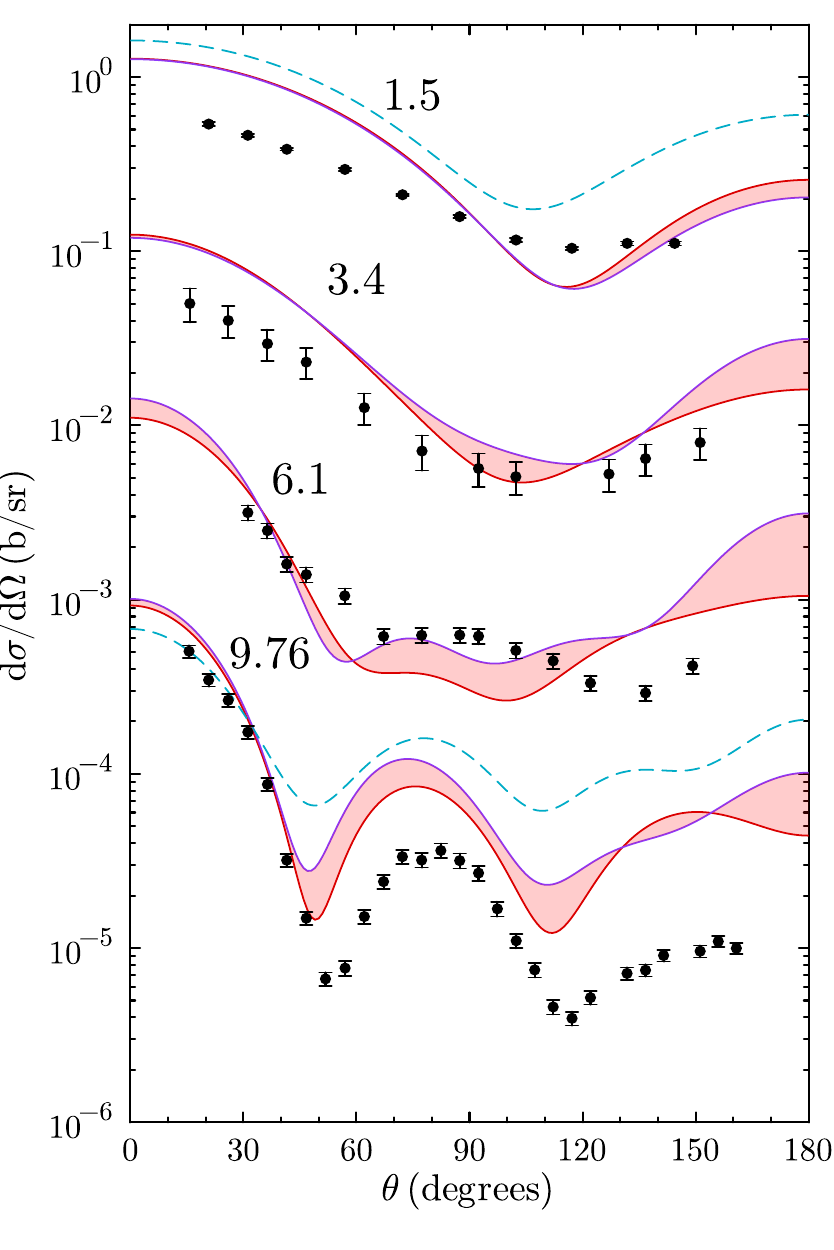}
    \caption{The differential cross section of elastic $\mg{24}+n$ scattering for four different neutron energies with respect to center--of--mass angle. The lab frame energies are shown in the figure in MeV.
        The lines and datapoints mean the same as in \autoref{fig:el}.
        The dashed lines at 1.5 and 9.76 MeV refer to the calculation with only the completion states. The experimental data are from \cite{Korzh:69,Schweitzer:78,Stewart:65,Virdis:81}.
        The cross sections for each energy are shifted down by a factor of 10 relative to the previous.}
    \label{fig:differential}
\end{figure}

The cross sections are calculated using two different odd nucleus bases. These basis states are
generated as eigenstates of the intrinsic $\exp (-i \pi \op{j}_x)$ operator with eigenvalue either $+i$ or $-i$
known as their signature quantum number. Since either basis forms a complete set of many-body states after
angular momentum projection \cite{Bally:21}, comparing the two can give an indication of the convergence of
the cross section calculation.
Additionally, cross sections were calculated using only the completion states by setting $s^x_{i,\alpha}=0$ in (\ref{eq:completion1},\ref{eq:completion2}) and are presented in \autoref{fig:el} and \ref{fig:differential}.

In \autoref{fig:total} the integrated neutron total cross sections are shown for energies from $E = \MeV{0.5}$ to $\MeV{13}$, together with the result of Koning--Delaroche optical potential \cite{Koning:03}.
We can see that below 10 MeV the calculation overestimates the cross section.
Due to the choice of $\eta$, resonances are wider than what experiments show at energies below $\MeV{5}$.

The elastic scattering cross section is also calculated and
compared to experiment in \autoref{fig:el}.
We see that the calculated cross section slightly overestimates the cross section up to $\MeV{9}$,
and this difference increases for neutron energies above $\MeV{10}$.
This indicates that the method fails to find enough
states that contribute to the non--elastic channels above $\MeV{9}$,
and progressively more of the contribution comes from
the mean field through the completion.
The potential obtained using only completion states is a real potential and only the elastic scattering channel is available in this case, resulting in a sizeable overestimation of the elastic scattering cross section in \autoref{fig:el}. At higher energy the calculated GCM $\sigma_{E,T}$ will gradually reach the result with only completion above 20 MeV.

The angular differential cross sections were calculated for several projectile energies, as shown in \autoref{fig:differential}, along with the corresponding experimental values. It is important to note that the solution to the scattering problem represents only one component of the experimentally measured cross section. When a neutron is absorbed, forming an excited $\mg{25}$ nucleus that subsequently decays by emitting another neutron of the same energy, this compound nucleus reaction contributes to the measured cross section and is indistinguishable from elastic scattering. Since the compound nucleus has time to thermalize, its emission is approximately isotropic, contributing to the differential cross section with little angular dependence. This effect is expected to be more significant at lower energies and around the minima.

To more accurately compare with experimental data, in \autoref{fig:differential} we include the compound nucleus contribution calculated using the semi-microscopic optical model of \cite{Bauge:01,Koning:23} (known as JLM), which is based on properties of nuclear matter and it is phenomenologically adjusted to reproduce scattering.
Developing a compound nucleus model for a general non--local potential is particularly challenging and beyond the scope of the present study, hence we use a local optical model as in \cite{Sargsyan:24}.
This additional component is necessary because of the use of scattering theory \eqref{eq:scattering} to describe elastic scattering, not due to our particular microscopic approach constructing the optical potential.

In conclusion, an energy weighted sum rule for spectroscopic amplitudes is introduced, interpreted as a generalized mean field, which bridges the gap between nuclear structure and scattering that arise due to the necessary truncations done in many--body methods.
The results demonstrate the viability of this new method, and it enables calculating scattering observables from a wide range of many--body methods and interactions.
In addition, the results suggest the lack of absorption observed when using a microscopically generated optical potential, that was linked to configurations beyond particle--hole excitation in \cite{Idini:19}, is partially taken into account through the collective degrees of freedom when using GCM.
This shows that GCM can capture necessary low energy correlations needed to describe low energy neutron scattering with a microscopic method using multiple reference states
,
though further research is required to better understand the resonance widths of GCM states, important in the description of both elastic and nonelastic scattering.
A need for compound scattering implementations for general non--local potentials was also identified.
This work extends microscopic reaction approaches to study deformed, heavy, and exotic nuclei, representing a significant step towards a unified model of nuclear structure and reaction.
The flexibility of this method will allow for further improvements through the use of new interactions, functionals, collective coordinates, and many--body expansions, and enables it to be used for the whole nuclear chart.

\section*{Acknowledgements}
This work has been supported by the Swedish Research Council (Vetenskapsr{\aa}det) VR~2020-03721, Knut and Alice Wallenberg foundation (KAW~2015.0021), Crafoord fundation, and Krapperup fundation. Computing was enabled by resources provided by the National Academic Infrastructure for Supercomputing in Sweden (NAISS), partially funded by the Swedish Research Council through grant agreement no. 2022-06725, and The Centre for Scientific and Technical Computing at Lund University (LUNARC).


\bibliographystyle{apsrev4-2}
\bibliography{sources.bib,nuclear.bib}

\end{document}


\renewcommand{\figurename}{Fig.}
\renewcommand{\figureautorefname}{Fig.}

\title{Supplementary: Nuclear cross sections from low-energy interactions}



\maketitle

\section{French--McFarlane sum rule}
Due to the relation,
\begin{equation}
\sum_\alpha \braopket{\Phi_0}{\hconj{a}_\alpha a_\alpha}{\Phi_0} = N,
\label{eq:1}
\end{equation}
where $N$ is the number of particles of a specific fermion species (proton or neutron) and $\alpha$ is an index spanning its single--particle Hilbert space, we obtain the French--McFarlane sum rule. This relates the spectroscopic factors of the $A-1$ states to the number of particles in the even--even ground state: $\sum_{i,\alpha} \left|s^-_{i,\alpha}\right|^2 = N$. However, since the states found by the GCM procedure will lack spectroscopic strength, the number of particles will be underestimated and the missing contribution must be added through the completion procedure. 

To construct the Green's function we choose to interpret completion amplitudes below a certain energy as occupied, choosing this energy to fulfill the French--McFarlane sum rule. But in the case of open--shell nuclei, there will still be one degenerate shell $I$ with energy $\epsilon_I$, which would give too many particles if occupied, but too few if not,
\begin{align}
    \sum_{i,\alpha} \left|s^-_{i,\alpha}\right|^2 + \sum_{\substack{k,\alpha \\\epsilon_k<\epsilon_I}} \left|c_{k,\alpha}\right|^2 & < N,
       \\
    \sum_{i,\alpha} \left|s^-_{i,\alpha}\right|^2 + \sum_{\substack{k,\alpha \\\epsilon_k<\epsilon_I}} \left|c_{k,\alpha}\right|^2 & + \sum_{\substack{k,\alpha \\\epsilon_k=\epsilon_I}}\left|c_{k,\alpha}\right|^2 > N.
\end{align}

Without affecting the two sum rules of Eqs. (5,6) in the paper, we can divide this shell $I$ in two, $I^-$ and $I^+$, by taking $\epsilon_{I^-} = \epsilon_{I^+} = \epsilon_I$, $c_{I^-,\alpha} = \sqrt{K} c_{I,\alpha}$ and $c_{I^+,\alpha} = \sqrt{1-K} c_{I,\alpha}$. We can then interpret the shell $I^-$ as occupied and $I^+$ as unoccupied. The particle number sum rule then becomes
\begin{equation}
    \sum_{i,\alpha} \left|s^-_{i,\alpha}\right|^2 + \sum_{\substack{k,\alpha\\\epsilon_k<\epsilon_I}} \left|c_{k,\alpha}\right|^2 + K \sum_{\substack{k,\alpha \\\epsilon_k=\epsilon_I}}\left|c_{k,\alpha}\right|^2 = N,
\end{equation}
which can be solved for $K$, corresponding to the fraction of $I$ which is occupied.
The energy $\epsilon_I$ then acts as the Fermi energy instead of the choice of $E_F$ described in the paper, and the shell $I$ is partially filled.
When constructing the Green's function, we then get $\bar{\epsilon}_i\of{\eta} = \epsilon_i + i\eta$ for $\epsilon_i < \epsilon_I$ and $i=I^-$, and $\bar{\epsilon}_i\of{\eta} = \epsilon_i - i\eta$ for $\epsilon_i > \epsilon_I$ and $i=I^+$.

\section{Calculation parameters}

The parameters of the many-body framework used are described in detail in \cite{Ljungberg:22}. In this work,
a spherical DFT calculation was preformed in a harmonic oscillator basis with 17 major shells using the SLy4 functional.
The lowest $586$ lowest energy single--particle wavefunctions (which correspond to approximately as many states as there would be in 11 major shells) was then
used to define a spherical single--particle basis and the effective Hamiltonian used in the GCM calculation.
We then sampled the collective coordinate space with 275 states. Keeping states below $\MeV{25}$ resulted in 179 many--body basis states forming the multiple reference states of the GCM calculation. The choice of $\MeV{25}$ excitation energy reflects the energy range considered, with the scattering projectile probing states around its energy plus separation energy.
The states were projected with 10 particle number projection points for protons and 10 for neutrons,
and $(6,12,24)$ angular momentum projection points for the three projection angles. Temperature parameter generating the particle--hole excitation was $\tilde b=0.45$ that was found in \cite{Ljungberg:22} to optimize convergence of excitations.
The odd--even basis consisted of one quasiparticle excitation for each even--even basis state. The quasiparticle to excite was chosen randomly among the 10 quasiparticles with the lowest mean-field energy of a given signature and $J\pi$. Quasiparticle states that correspond to the wrong particle number are excluded, i.e. hole states when calculating the solution for $A+1$.

All these parameters have been chosen to optimize convergence, measured as the variance between $\pm i$ signatures on either the spectrum of odd nuclei, or scattering observables. The previous study of \cite{Ljungberg:22} found optimal convergence parameters for $^{24}$Mg.

In our calculations, the optical potential $V^{J\pi}_{\alpha,\beta}\of{E}$ is first expressed in a finite harmonic oscillator basis. When transforming to momentum basis, a smoothing factor
\begin{equation}
    \sigma_n = \frac{1-\exp\of{-\left(\alpha \frac{n-N^{J\pi}}{N^{J\pi}}\right)^2}}{1-\exp\of{-\alpha^2}}
\end{equation}
is applied to the matrix elements, where $\alpha$ is a dimensionless parameter as described in \cite{Revai:85},
giving the smoothed potential $\tilde{V}^{J\pi}$ as
\begin{equation}
    \tilde{V}^{J\pi}_{n m} = \sigma_n V^{J\pi}_{n m} \sigma_m,
\end{equation}
which is then used in place of $V^{J\pi}$ in the scattering.

This smoothing factor improves convergence by
decreasing the effect of the basis truncation
\cite{Revai:85,shirokov:2021,du:2022},
and is similar to the introduction of the Lanczos factors
that were originally devised to reduce the Gibbs phenomenon
in Fourier analysis.
We set $\alpha = 5$, which was verified by calculating scattering cross sections using only the single--particle part $\op{\Gamma}$ as potential, giving the same cross section from a calculation with 17 major shells as one with 50 major shells.




\bibliographystyle{apsrev4-2}
\bibliography{sources.bib,nuclear.bib}